\documentclass{article}
\usepackage{spconf,amsmath,graphicx}
\usepackage{booktabs}
\usepackage{multirow}
\usepackage{subfig}
\title{Inter-KD: Intermediate Knowledge Distillation for CTC-Based Automatic Speech Recognition}
\name{Ji Won Yoon$^1$, Beom Jun Woo$^1$, Sunghwan Ahn$^1$, Hyeonseung Lee$^1$, and Nam Soo Kim$^1$}
\address{$^1$Department of ECE and INMC, Seoul National University, Seoul,
Republic of Korea\\
}
\begin{document}
%\ninept
%
\copyrightnotice{978-1-6654-7189-3/22/\$31.00~\copyright2023 IEEE}
\maketitle
\begin{abstract}
Recently, the advance in deep learning has brought a considerable improvement in the end-to-end speech recognition field, simplifying the traditional pipeline while producing promising results.
Among the end-to-end models, the connectionist temporal classification (CTC)-based model has attracted research interest due to its non-autoregressive nature.
However, such CTC models require a heavy computational cost to achieve outstanding performance.
To mitigate the computational burden, we propose a simple yet effective knowledge distillation (KD) for the CTC framework, namely Inter-KD, that additionally transfers the teacher's knowledge to the intermediate CTC layers of the student network.
From the experimental results on the LibriSpeech, we verify that the Inter-KD shows better achievements compared to the conventional KD methods.
Without using any language model (LM) and data augmentation, Inter-KD improves the word error rate (WER) performance from 8.85 \% to 6.30 \% on the test-clean.
% Also, we show that the proposed KD can be applied to the early exit framework that can accelerate the model's inference speed.

\end{abstract}
\begin{keywords}
Speech recognition, connectionist temporal classification, teacher-student learning, knowledge distillation
\end{keywords}
\section{Introduction}
\label{sec:intro}
In recent years, there has been remarkable progress in end-to-end speech recognition that directly converts an input speech into the corresponding text without any prior alignment information.
Compared with the traditional deep neural network (DNN)-hidden Markov model (HMM) hybrid systems, the end-to-end framework simplifies the overall pipeline while achieving better performance.

Among the various types of end-to-end models for speech recognition, connectionist temporal classification (CTC) \cite{graves-et-al:scheme} has attracted increasing interest due to its non-autoregressive (NAR) nature.
The NAR model requires $M$($\ll N$) iterations when producing an $N$-length target sequence.
On the other hand, the autoregressive (AR) model costs $N$ iterations, indicating that the NAR framework enables a significant inference speedup over the AR one.

\begin{figure}[t]
    \centering
        \subfloat[Previous KD approach \label{intro1}]{\includegraphics[width=6.5cm]{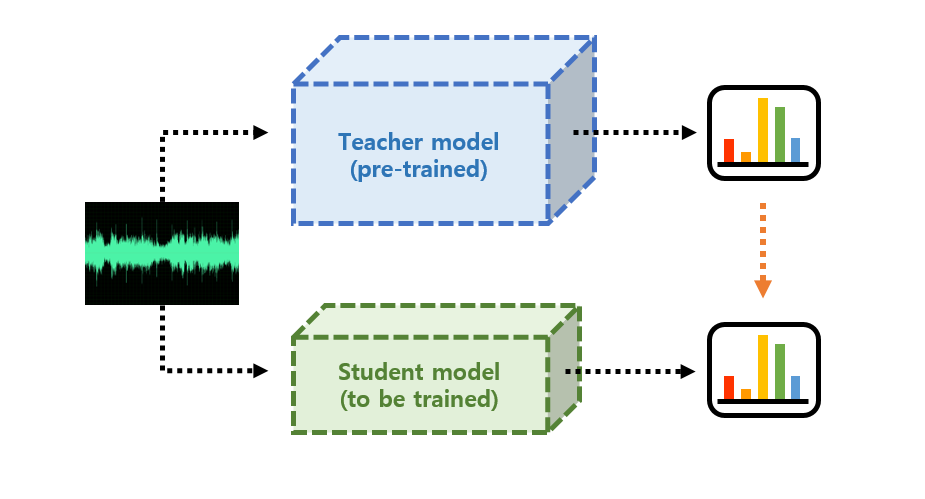}}
        \qquad %
        \subfloat[Proposed KD approach \label{intro2}] {\includegraphics[width=6.5cm]{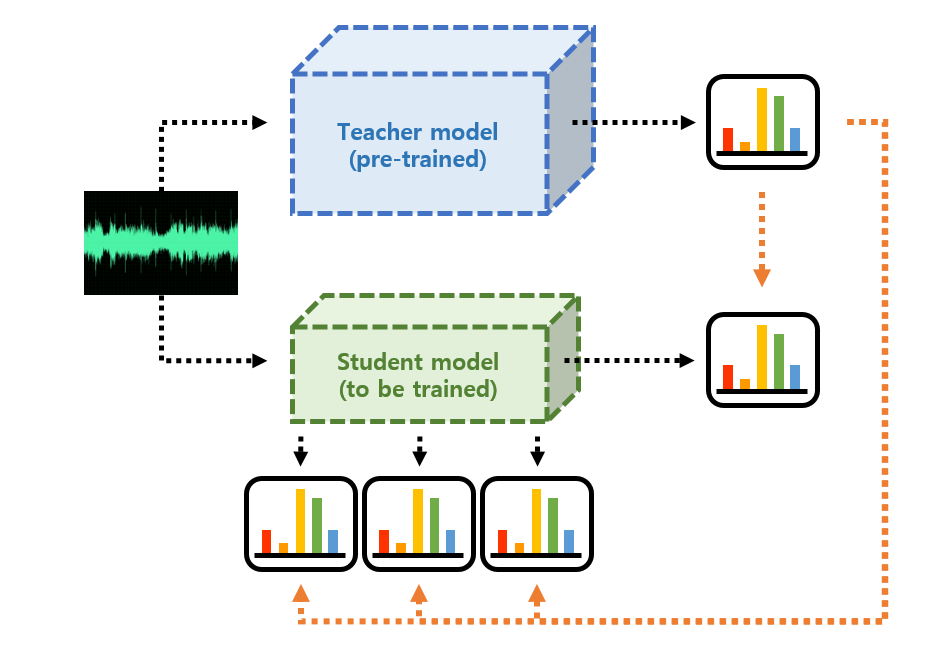}}
    \caption{Conceptual diagram of Inter-KD compared to the conventional KD. The orange line represents KD with the teacher's output-level knowledge. Different from the conventional KD, the proposed method uses multiple intermediate CTC layers for KD.}
  \label{intro}
\end{figure}

However, existing CTC-based models require high computational cost and long training time to achieve promising results.
For their practical deployment in resource-limited settings, there have been continuous efforts to apply knowledge distillation (KD), an effective technique for model compression, to CTC models.
The main idea of KD is to transfer the knowledge from a large and powerful teacher model to a small student model.
In the speech recognition task, conventional KD methods typically consider the teacher network's output-level knowledge (sentence prediction, softmax prediction, etc.) for distillation.
By minimizing the distance between the predictions of the teacher and the student, the distilled student can achieve better performance than the baseline trained only with the target label.

In order to train the student more effectively, in this paper, we propose a novel KD framework for CTC models, namely Inter-KD.
As depicted in Fig. \ref{intro}, we design the student model's architecture using multiple intermediate  CTC layers.
Each CTC layer is trained with the output-level knowledge of the teacher in conjunction with the CTC loss.
The proposed architecture of the student is similar to a deeply supervised net \cite{deeply} and an intermediate CTC \cite{inter-ctc}, where both methods add supervision to the hidden layers for performance improvement.
The main difference with previous deeply supervised schemes is that Inter-KD trains the intermediate CTC layers with KD instead of using only ground-truth labels.
Since the auxiliary CTC layers can be ignored during the inference, a small additional computational load is required only for training.

From the experimental results on the LibriSpeech \cite{panayotov-et-al:scheme} dataset, it is confirmed that Inter-KD shows better performance than the conventional KD methods.
For test-clean dataset, Inter-KD improves the word error rate (WER) performance of the student from 8.85 \% to 6.30 \% without using any language model (LM) and data augmentation, achieving relative error rate reduction (RERR) 28.81 \%.
We conduct additional analysis to further check the effect of KD with the intermediate CTC layers.

Our main contributions can be summarized as follows:
\begin{itemize}
    \item We introduce a new KD framework for CTC models, namely Inter-KD.
    In the proposed scheme, we newly design the architecture of the student by attaching multiple intermediate CTC layers in the middle of the network.
    The student can be trained more effectively by transferring the teacher's knowledge to the intermediate CTC layers.
    \item According to the experimental results on the LibriSpeech dataset, we verify the effectiveness of the Inter-KD. 
    When transferring the output-level knowledge of the teacher, Inter-KD yields better performance than other previous KD methods.

    % \item We also show that the Inter-KD can be applied to the early exit framework, which can dynamically adjust the inference latency-performance trade-off.

\end{itemize}

\section{Related Work}
\label{sec:related_work}
\subsection{Connectionist temporal classification}
For given source input $x_{1:T}$ and target label $y_{1:N}$, the CTC framework \cite{graves-et-al:scheme} can directly convert $x$ into $y$ by using the additional token ``blank".
Unlike the traditional hybrid system, the predefined alignment knowledge is not required.
CTC considers all possible alignments compatible with $y$ to compute the conditional probability of $y$.
When training the CTC model, we minimize the following objective:
\begin{equation}
   -\log p(y|x) = -\log \sum_{a\in\beta^{-1}(y)}p(a|x)
   \label{ctc1}
\end{equation}
where $\beta$ denotes a many-to-one mapping function for CTC, and $a$ represents the intermediate alignments, which include the blank token. $\beta^{-1}(y)$ returns the possible set of alignments.

Even though the CTC framework provides efficient decoding, there is a strong conditional independence assumption between the output tokens, resulting in relatively poor performance compared to the AR models.
Recently, there have been some attempts to close the gap between the CTC and AR models. Chan \textit{et al.} \cite{imputer} and Higuchi \textit{et al.} \cite{mask-ctc} used the additional network to refine the initial output from the CTC.
Majumdar \textit{et al.} \cite{citrinet} proposed an improved CTC-based architecture that combines a QuartzNet \cite{quartznet:scheme} with the squeeze and excitation \cite{sqeeze}.
Lee and Watanabe \cite{inter-ctc} introduced an intermediate CTC loss, which uses the intermediate layer in the encoder network and its corresponding CTC loss to improve the performance of the model.
\subsection{Knowledge distillation}
% KD is one of the most effective methods for model compression.
Hinton \textit{et al.} \cite{hinton_kd-et-al:scheme} first proposed the concept of KD, which transfers knowledge by minimizing Kullback-Leibler (KL)-divergence between the predictions of the teacher and the student.
Since the large and powerful teacher model is utilized to guide the training of the small student model, the student can produce better performance compared with the case when it is solely trained with the ground-truth labels.
With steadily increasing interest in on-device speech recognition, there have been several efforts to develop KD for speech recognition models.
For the DNN-HMM hybrid system, previous KD studies typically trained the student by minimizing cross-entropy (CE) loss between the posterior probability of the teacher and the student \cite{firstasr:scheme,chebotar-et-al:scheme,watanabe-et-al:scheme,lu-et-al:scheme,fukuda-et-al:scheme, blending:scheme}.
However, it is challenging to train CTC-based speech recognition models with the same CE criteria.
According to the previous studies \cite{senior-et-al:scheme,takashima-et-al:scheme, takashima-et-al2:scheme}, applying the CE-based KD technique to the CTC-based models can worsen the performance compared to the baseline model trained only with the target label.
To cover this issue, Takashima \textit{et al.} \cite{takashima-et-al:scheme} attempted to apply sequence-level KD \cite{sequence-level-kd} to the CTC model. 
Kurata and Audhkhasi \cite{kurata2-et-al:scheme,kurata-et-al:scheme} suggested the KD framework that can train the low-latency student model with the knowledge of the high-latency teacher model and also proposed guided CTC training that distills the spike timings from the teacher.
Yoon \textit{et al.} \cite{tutornet:scheme} introduced softmax-level KD (SKD) that uses $l_2$ loss instead of KL divergence when distilling frame-level posterior of the CTC-based teacher model.

\section{Proposed Method}
\label{sec:proposed}
\begin{figure*}[t]
    \centering
        \includegraphics[height=8cm]{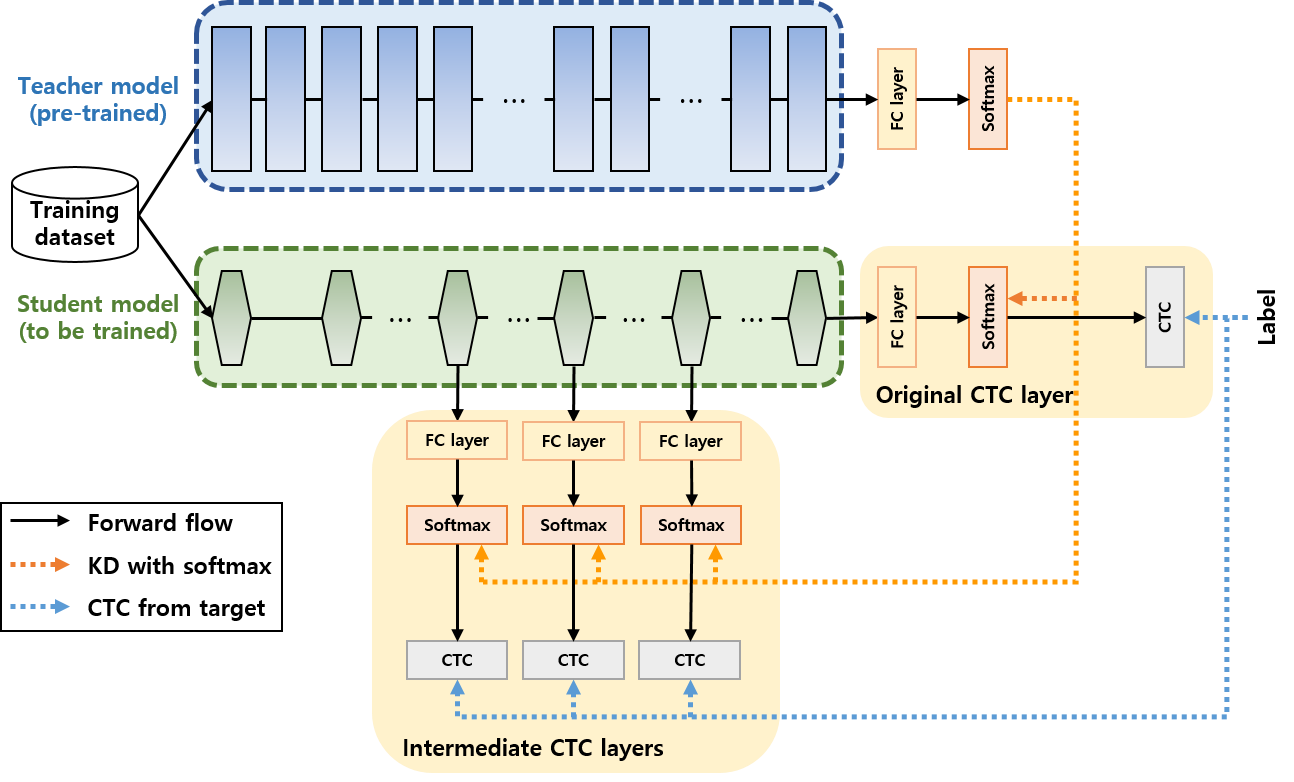}
    \caption{An overview of Inter-KD for CTC models.}
  \label{proposed}
\end{figure*}
\subsection{Student model architecture}
As shown in Fig. \ref{proposed}, we newly design the architecture of the student. 
Different from the conventional student model, multiple CTC layers are added to the intermediate layers of the student.
In our experiments, additional CTC layers are attached to three layers: $18^{th}$, $24^{th}$, and $30^{th}$ layers of the student, whose network is composed of 33 depthwise separable convolutional layers. 
For the convenience of notation, we let ``intermediate CTC layer" denotes the auxiliary CTC layer in the middle of the student, and ``original CTC layer" denotes the last CTC layer of the student. Each CTC layer is combined with a fully-connected layer and the softmax function.
Note that the fully-connected layers in intermediate CTC are not shared with the original CTC’s fully-connected layer.
The proposed architecture of the student is similar to some deep supervision-based networks \cite{deeply, inter-ctc} that add supervision to the hidden layers.
The main difference is that the intermediate CTC layers are trained with KD instead of using only ground-truth labels.

\subsection{Intermediate CTC layer}
Intermediate CTC layers are located in the middle of the student network, as depicted in Fig. \ref{proposed}.
These layers are only used in the training procedure and can be removed during the inference, so a small additional computational load is required only for training.
When training the student, the intermediate CTC layers are trained via KD in conjunction with CTC loss.
The CTC loss function of $i^{th}$ intermediate CTC layer is given as
\begin{equation}
   \mathcal{L}_{\text{InterCTC}}^{i} = CTC_{\text{loss}}(y, g_{i}(x))
   \label{inter_ctc}
\end{equation}
where $x$, $y$, $g_{i}(x)$, and $CTC_{\text{loss}}$ represent the source input, the corresponding label, the softmax output of the $i^{th}$ intermediate CTC layer, and the CTC loss, respectively.
Regardless of the layer position, the conventional deep supervised learning \cite{deeply, inter-ctc} added supervision to the hidden layers with the same target. Since we
followed the conventional framework, we used the same target labels for each intermediate layer.

The second loss source is the KD loss function. As for the KD technique, we adopt the SKD \cite{tutornet:scheme} due to its effective improvement for CTC models. 
The SKD applies $l_{2}$ loss for transferring the knowledge.
Therefore, the KD loss function for $i^{th}$ intermediate CTC layer can be computed as
\begin{equation}
   \mathcal{L}_{\text{InterKD}}^{i} = ||f_{tea}(x)-g_{i}(x))||^{2}_{2}
   \label{inter_kd}
\end{equation}
where $f_{tea}(x)$ denotes the softmax prediction of the teacher.

\subsection{Original CTC layer}
The original CTC layer is attached to the last layer of the student model. 
We use both KD and CTC losses for training the original CTC layer. 
Firstly, the CTC loss function of the original CTC layer is as follows:
\begin{equation}
   \mathcal{L}_{\text{OrigCTC}} = CTC_{\text{loss}}(y, f_{stu}(x))
   \label{orig_ctc}
\end{equation}
where $ f_{stu}(x)$ denotes the softmax value of the original CTC layer.
$\mathcal{L}_{\text{OrigCTC}}$ loss is exactly same as the vanilla CTC training.
The KD loss for the original CTC layer is formulated as
\begin{equation}
   \mathcal{L}_{\text{OrigKD}} = ||f_{tea}(x)-f_{stu}(x))||^{2}_{2}.
   \label{orig_kd}
\end{equation}

\begin{table*}[t]
\centering
\caption{Comparison of WER (\%) and RERR (\%) on LibriSpeech using greedy decoding.~``Ours" denotes Inter-KD with $K=3$. The best result is in bold.}
\label{totalwer_nolm}
\begin{tabular}{@{}lcrrrrrrrr@{}}
\toprule
 && \multicolumn{4}{c}{WER (\%) w/o LM} & \multicolumn{4}{c}{RERR (\%) w/o LM} \\ \cmidrule(l){3-10} 
 \hspace{15mm}Model &Params.& \multicolumn{2}{c}{dev} & \multicolumn{2}{c}{test} & \multicolumn{2}{c}{dev} & \multicolumn{2}{c}{test} \\ \cmidrule(l){3-10} 
& & clean & \multicolumn{1}{c}{other} & clean & \multicolumn{1}{c}{other} & clean & \multicolumn{1}{c}{other} & clean & \multicolumn{1}{c}{other}  \\ \midrule
Teacher: Jasper DR  & 332.63 M & 3.61 & 11.37 & 3.77 & 11.08 & -\hspace{3mm} & -\hspace{3mm} & -\hspace{3mm} & -\hspace{3mm} \\
\midrule
Student: Jasper Mini&  & 8.66 & 23.28 & 8.85 & 24.26 &  -\hspace{3mm} & -\hspace{3mm} & -\hspace{3mm} & -\hspace{3mm} \\
\hspace{3mm}+ Sequence-level KD \cite{takashima-et-al:scheme} && 8.96 & 23.73 & 9.10 & 24.81 & -3.46 & -1.93 & -2.82 & -2.27\\
\hspace{3mm}+ Guided CTC training \cite{kurata-et-al:scheme} & 8.19 M & 7.81 & 21.93 & 8.29 & 22.49 &9.82 & 5.80 & 6.33 & 7.30\\
\hspace{3mm}+ SKD \cite{tutornet:scheme}  && 7.63 & 21.36 & 7.81 & 22.41 & 11.89 & 8.25 & 11.75 & 7.63\\
\hspace{3mm}\textbf{+ Ours} && \textbf{6.24} & \textbf{18.82} & \textbf{6.30} & \textbf{19.49} &\textbf{27.94}&\textbf{19.16}&\textbf{28.81}&\textbf{19.66} \\
\bottomrule
\end{tabular}
\end{table*}

\subsection{Training}
In the Inter-KD framework, there are two kinds of training losses to improve the performance of the student.
\begin{itemize}
    \item Loss 1: CTC loss from target labels. The supervision of CTC is added not only to the original CTC layer, but also to the intermediate CTC layers.
    \item Loss 2: KD loss using the teacher model's softmax prediction. All CTC layers are trained with the knowledge of the teacher.
\end{itemize}
When there are $K$ intermediate CTC layers in the student model, the CTC loss function $\mathcal{L}_{\text{CTC}}$ for Inter-KD is given as
\begin{equation}
   \mathcal{L}_{\text{CTC}} = \mathcal{L}_{\text{OrigCTC}} + \sum_{i=1}^K \mathcal{L}_{\text{InterCTC}}^{i}.
   \label{tot_ctc}
\end{equation}
The KD loss between the student and the teacher is as follows:
\begin{equation}
   \mathcal{L}_{\text{KD}} = \mathcal{L}_{\text{OrigKD}} + \sum_{i=1}^K \mathcal{L}_{\text{InterKD}}^{i}.
   \label{tot_kd}
\end{equation}
Thus, the final objective function $\mathcal{L}_{\text{Total}}$ for Inter-KD is given as
\begin{equation}
   \mathcal{L}_{\text{Total}} = \mathcal{L}_{\text{CTC}} + \lambda \cdot \mathcal{L}_{\text{KD}}
   \label{tot_kd}
\end{equation}
where $\lambda$ is a tunable parameter to balance $\mathcal{L}_{\text{CTC}}$ and $\mathcal{L}_{\text{KD}}$.

\subsection{Inference}
The intermediate CTC layers are unaffected during the whole inference procedure.
Only the original CTC layer is used to generate the final prediction of the student.
% However, we can additionally adopt the intermediate layers to perform the early-exit inference, accelerating the inference speed. It will be discussed in Section .

\section{Experimental Settings}
\label{sec:exp_setting}
\subsection{Dataset}
We evaluated the word error rate (WER) performance on LibriSpeech \cite{panayotov-et-al:scheme} dataset.
In the training phase, ``train-clean-100", ``train-clean-360", and ``train-other-500" were applied. We used ``dev-clean", ``dev-other", ``test-clean", and ``test-other" for evaluation.

\subsection{Performance metrics}
For the performance comparison, we measured WER and relative error rate reduction (RERR).
WER is a widely-used metric to quantify the performance of the speech recognition model and RERR is a standard metric to measure the WER improvement compared to the baseline.

\subsection{Model configurations}
In our experiments, we adopted Jasper DR \cite{li-et-al:scheme} and Jasper Mini as the teacher and the student models, respectively.
Both CTC models had the same label set, which included a total of 29 character labels.
For model training and inference, We utilized the OpenSeq2Seq toolkit \cite{openseq2seq}.
For the Jasper DR teacher, we used the pre-trained model checkpoint provided by the OpenSeq2Seq.
The Jasper Mini student model consists of 33 depthwise separable 1D convolutional layers.

\subsection{Implementation details}
We trained the student with 50 epochs for CTC training with KD, and three Titan V GPUs (each with 12GB memory) were used for training.
NovoGrad optimizer \cite{ginsburg-et-al:scheme} was adopted for training the student, where the initial learning rate was set to 0.02.
As aforementioned in Section \ref{sec:proposed}, we added intermediate CTC layers to three layers: $18^{th}$, $24^{th}$, and $30^{th}$ layers of the Jasper Mini.
The tunable parameter in the Equation \ref{tot_kd} was set to 0.25.
When decoding with LM, we used 4-gram KenLM \cite{Heafield2011KenLMFA}, where the beam width was 256.

\subsection{Conventional distillation methods for comparison}
Since the proposed framework transferred the softmax prediction of the teacher, we mainly compared the Inter-KD with the conventional KD methods that considered the teacher network's output-level knowledge (sentence prediction, softmax prediction, etc.).
For performance comparison, we applied three conventional KD techniques for CTC models, including sequence-level KD \cite{takashima-et-al:scheme}, guided CTC training \cite{kurata-et-al:scheme}, and SKD \cite{tutornet:scheme}.

\begin{table*}[t]
\centering
\caption{Comparison of WER (\%) and RERR (\%) using the 4-gram LM. ~``Ours" denotes Inter-KD with $K=3$. The best result is in bold.}
\label{totalwer_lm}
\begin{tabular}{@{}lcrrrrrrrr@{}}
\toprule
 && \multicolumn{4}{c}{WER (\%) w/ LM} & \multicolumn{4}{c}{RERR (\%) w/ LM} \\ \cmidrule(l){3-10} 
 \hspace{15mm}Model &Params.& \multicolumn{2}{c}{dev} & \multicolumn{2}{c}{test} & \multicolumn{2}{c}{dev} & \multicolumn{2}{c}{test} \\ \cmidrule(l){3-10} 
& & clean & \multicolumn{1}{c}{other} & clean & \multicolumn{1}{c}{other} & clean & \multicolumn{1}{c}{other} & clean & \multicolumn{1}{c}{other}  \\ \midrule
Teacher: Jasper DR  & 332.63 M & 3.04 & 9.52 & 3.69 & 9.38 & -\hspace{3mm} & -\hspace{3mm} & -\hspace{3mm} & -\hspace{3mm} \\
\midrule
Student: Jasper Mini&  &  4.83 & 15.53 & 5.24 & 16.40& -\hspace{3mm} & -\hspace{3mm} & -\hspace{3mm} & -\hspace{3mm} \\
\hspace{3mm}+ Sequence-level KD \cite{takashima-et-al:scheme} && 5.16 & 15.54 & 5.48 & 16.91 & -6.83 & -0.06 & -4.58 & -3.11\\
\hspace{3mm}+ Guided CTC training \cite{kurata-et-al:scheme} & 8.19 M & 5.17 & 15.94 & 5.58 & 16.85 & -7.04 & -2.64 & -6.49 & -2.74\\
\hspace{3mm}+ SKD \cite{tutornet:scheme} && 4.77 & 15.01 & 5.26 &15.96 & 1.24 & 3.35 & -0.38 & 2.68\\
\hspace{3mm}\textbf{+ Ours} && \textbf{4.61} & \textbf{14.55} & \textbf{4.99} & \textbf{15.19} & \textbf{4.55} & \textbf{6.31}& \textbf{4.77} & \textbf{7.38}\\
\bottomrule
\end{tabular}
\end{table*}

\section{Experimental Results}
\label{sec:exp_result}
\subsection{Performance comparison}
Table \ref{totalwer_nolm} shows the WER and RERR results with greedy decoding, comparing the performance improvement of the Inter-KD with conventional KD techniques.
From the results, it is verified that the proposed KD method considerably improved the WER performance of the student compared to other KD methods.
The distilled student using Inter-KD achieved WER 6.24 \% and WER 6.30 \% on dev-clean and test-clean, corresponding to RERR 27.94 \% and RERR 28.81 \%.
In the case of test-other, Inter-KD gave WER 19.49 \% and RERR 19.66 \%.
Among the conventional KD approaches, SKD yielded better achievements than sequence-level KD and guided CTC training.
Still, the best WER performance was obtained when applying the Inter-KD to the student model.

Also, we conducted experiments based on the LM decoding. As presented in Table \ref{totalwer_lm}, applying LM was more challenging than greedy decoding. 
Sequence-level KD and guided CTC training had a negative RERR value for all configurations, and SKD had a little improvements compared to the case when utilizing greedy decoding.
Compared to the previous results in Table \ref{totalwer_nolm}, the conventional KD techniques did not perform well with the LM decoding.
However, it is confirmed that Inter-KD gave significant improvements in all configurations, even when decoding with LM.
The proposed KD achieved WER 4.99 \% and WER 15.19 \% on test-clean and test-other, providing RERR 4.77 \% and RERR 7.38 \%, respectively.

\subsection{Analysis}
\begin{table}[t]
\centering
\caption{WER (\%) and RERR (\%) on LibriSpeech dev-clean using greedy decoding. Different number of intermediate CTC layers (=$K$) were set from 1 to 3. RERR measured the WER improvement compared to the baseline, where the student baseline provided WER 8.66 \% on dev-clean.}
\label{layer_ana}
\begin{tabular}{@{}ccrr@{}}
\toprule
 {\begin{tabular}[c]{@{}c@{}}\# of intermediate\\CTC layers\end{tabular}} & KD &  WER (\%) & RERR (\%) \\ 
 \midrule
 \multirow{2}{*}{K=3} & O & \textbf{6.24} \hspace{3mm} & \textbf{27.94} \hspace{3mm} \\
& X& 6.60 \hspace{3mm} & 23.79 \hspace{3mm} \\
 \midrule
 \multirow{2}{*}{K=2}& O & 6.29 \hspace{3mm} & 27.37 \hspace{3mm} \\
& X & 7.37 \hspace{3mm} & 14.90 \hspace{3mm} \\
 \midrule
  \multirow{2}{*}{K=1}& O & 6.42 \hspace{3mm} & 25.87 \hspace{3mm} \\
 & X & 8.20 \hspace{3mm} & 5.31 \hspace{3mm} \\

 \bottomrule
\end{tabular}
\end{table}
\begin{table}[t]
\centering
\caption{WER (\%) on LibriSpeech when greedy decoding was applied. $K$ was set to 3.}
\label{ee}
\begin{tabular}{@{}crrrr@{}}
\toprule
 & \multicolumn{4}{c}{WER (\%)} \\ \cmidrule(l){2-5} 
 {\begin{tabular}[c]{@{}c@{}}Intermediate CTC\\layer index\end{tabular}}  & \multicolumn{2}{c}{dev} & \multicolumn{2}{c}{test} \\ \cmidrule(l){2-5} 
 & clean & \multicolumn{1}{c}{other} & clean & \multicolumn{1}{c}{other}  \\ \midrule
 1 &11.91	&28.91&	12.03&	29.81\\
2 &7.36&	21.00 &	7.39&	21.76\\
3 &6.30&	18.97&	6.37&	19.48\\

 \bottomrule
\end{tabular}
\end{table}
\subsubsection{Effect of the number of intermediate CTC layers}
\label{effect_ana}
In addition to the performance comparison, we set different number of intermediate CTC layers to verify the impact of $K$, which denotes the number of intermediate CTC layers.
Table \ref{layer_ana} gives the WER and RERR results on dev-clean by setting different $K$ from 1 to 3.
When $K$ was set to 1, we attached one intermediate CTC layer to the 30$^{th}$ layer of the student model.
In the case of $K=2$, the intermediate CTC layers were added to 24$^{th}$ and 30$^{th}$ layers of the student.
The intermediate layers are attached to 18$^{th}$, 24$^{th}$, and 30$^{th}$ layers of the student when $K=3$.
From the results, we confirmed that, regardless of applying KD, the performance progressively improved as the $K$ increased from 1 to 3.

\subsubsection{Performance comparison with deep supervised learning}
The conventional deep supervised learning \cite{deeply}, such as intermediate CTC \cite{inter-ctc}, added supervision to the hidden layers.
Therefore, training the intermediate CTC layers without KD was similar to the deep supervised scheme.
In addition to the previous experimental results, we continued to conduct the performance comparison between Inter-KD and the deep supervised learning.
The results in Table \ref{layer_ana} show that Inter-KD achieved better improvements compared to the case without KD, where multiple intermediate CTC layers were trained only with the ground-truth.
With the setting of $K=1$, there was relatively little achievement without KD, providing RERR 5.31 \%.
We observed that the distilled student ($K=1$) using Inter-KD was improved considerably, yielding RERR 25.87 \%. 
In the case of $K=2$, the model provided WER 7.37 \% without KD, and the performance was further improved when applying Inter-KD (WER 6.29 \%).
We verified that applying KD for each intermediate CTC layer performed well for all configurations, including $K=1$, $K=2$, and $K=3$.
Our best improvement was obtained when applying $K=3$ with Inter-KD.

\subsubsection{Performance of each intermediate CTC layer}
Since we used multiple CTC intermediate layers that can produce the intermediate ASR prediction, we measured the WER performance of each intermediate CTC layer. Table \ref{ee} summarizes the results of each intermediate CTC layer when applying $K=3$. Layer indexes 1, 2, and 3 indicate each intermediate layer attached to the 18$^{th}$, 24$^{th}$, and 30$^{th}$ layers of the student model, respectively.
From the results, the intermediate CTC with index 1 achieved WER 12.03 \% and WER 29.81 \% on test-clean and test-other datasets, indicating a substantial WER degradation compared to other layer indexes.
In the case of index 3, we observed a minimal degradation for all configurations, compared with the previous results of Inter-KD in Table \ref{totalwer_nolm}. 
Interestingly, for the test-other, the WER performance of layer index 3 (WER 19.48 \%) was slightly better than that of the original Inter-KD (WER 19.49 \%).
This implies that intermediate CTC layers can produce confident ASR predictions via the Inter-KD.
If we properly use these intermediate outputs for the early exit framework, the ASR result can be returned early instead of finishing the whole inference procedure.
In other words, the Inter-KD gave more possibilities for accelerating the model's inference speed.

\section{Conclusion}
In this paper, we proposed a simple yet effective KD method for CTC models, namely Inter-KD.
In the proposed KD framework, we newly designed the architecture of the student by adding multiple intermediate CTC layers in the middle of the student network.
These intermediate layers were trained with the teacher's knowledge in conjunction with the ground-truth labels.
From the experimental results on LibriSpeech, it is confirmed that Inter-KD can effectively improve the WER performance of the student.
Also, the proposed KD achieved better improvements for all configurations compared to the conventional output-level KD methods.
The detailed analysis was performed for each intermediate CTC layer, and we gave the possibility that the proposed KD can be applied to the early exit framework, accelerating the model's inference speed.

\label{sec:conclusion}

% \section{ACKNOWLEDGMENTS}
% \label{sec:ack}
% This work was supported by Samsung Electronics Co., Ltd. (IO201211-08075-01)

\bibliographystyle{IEEEbib}
\bibliography{strings,refs}

\begin{thebibliography}{10}

\bibitem{graves-et-al:scheme}
A.~Graves, S.~Fern{\'a}ndez, F.~Gomez, and J.~Schmidhuber,
\newblock ``Connectionist temporal classification: labelling unsegmented
  sequence data with recurrent neural networks,''
\newblock in {\em Proc. ICML}, 2006, pp. 369--376.

\bibitem{deeply}
C.~Lee, S.~Xie, P.~Gallagher, Z.~Zhang, and Z.~Tu,
\newblock ``Deeply-supervised nets,''
\newblock in {\em Proc. AISTATS}, 2015.

\bibitem{inter-ctc}
J.~Lee and S.~Watanabe,
\newblock ``Intermediate loss regularization for ctc-based speech
  recognition,''
\newblock in {\em Proc. ICASSP}, 2021.

\bibitem{panayotov-et-al:scheme}
V.~Panayotov, G.~Chen, D.~Povey, and S.~Khudanpur,
\newblock ``Librispeech: an asr corpus based on public domain audio books,''
\newblock in {\em Proc. ICASSP}, 2015, pp. 5206--5210.

\bibitem{imputer}
W.~Chan, C.~Saharia, G.~Hinton, M.~Norouzi, and N.~Jaitly,
\newblock ``Imputer: sequence modelling via imputation and dynamic
  programming,''
\newblock in {\em Proc. ICML}, 2020.

\bibitem{mask-ctc}
Y.~Higuchi et~al.,
\newblock ``Mask ctc: non-autoregressive end-to-end asr with ctc and mask
  predict,''
\newblock in {\em Proc. INTERSPEECH}, 2020.

\bibitem{citrinet}
S.~Majumdar, J.~Balam, O~Hrinchuk, V.~Lavrukhin, V.~Noroozi, and B.~Ginsburg,
\newblock ``Citrinet: closing the gap between non-autoregressive and
  autoregressive end-to-end models for automatic speech recognition,''
\newblock {\em arXiv preprint arXiv:2104.01721v1}, 2021.

\bibitem{quartznet:scheme}
S.~Kriman, K.~Beliaev, B.~Ginsburg, J.~Huang, O.~Kuchaiev, V.~Lavrukhin,
  R.~Leary, J.~Li, and Y.~Zhang,
\newblock ``Quartznet: deep automatic speech recognition with 1d time-channel
  separable convolutions,''
\newblock {\em arXiv preprint arXiv:1910.10261}, 2019.

\bibitem{sqeeze}
J.~Hu, L.~Shen, and G.~Sun,
\newblock ``Squeeze-and-excitation networks,''
\newblock in {\em Proc. CVPR}, 2018.

\bibitem{hinton_kd-et-al:scheme}
G.~Hinton, O.~Vinyals, and J.~Dean,
\newblock ``Distilling the knowledge in a neural network,''
\newblock in {\em Proc. NIPS Workshop Deep Learn.}, 2014.

\bibitem{firstasr:scheme}
J.~Li, R.~Zhao, T.~J. Huang, and Y.~Gong,
\newblock ``Learning small-size dnn with output-distribution-based criteria,''
\newblock in {\em Proc. INTERSPEECH}, 2014.

\bibitem{chebotar-et-al:scheme}
Y.~Chebotar and A.~Waters,
\newblock ``Distilling knowledge from ensembles of neural networks for speech
  recognition,''
\newblock in {\em Proc. INTERSPEECH}, 2016, pp. 3439--3443.

\bibitem{watanabe-et-al:scheme}
S.~Watanabe, T.~Hori, J.~L.~Roux, and J.~R. Hershey,
\newblock ``Student-teacher network learning with enhanced features,''
\newblock in {\em Proc. ICASSP}, 2017, pp. 5275--5279.

\bibitem{lu-et-al:scheme}
L.~Lu, M.~Guo, and S.~Renals,
\newblock ``Knowledge distillation for small-footprint highway networks,''
\newblock in {\em Proc. ICASSP}, 2017, pp. 4820--4824.

\bibitem{fukuda-et-al:scheme}
T.~Fukuda, M.~Suzuki, G.~Kurata, S.~Thomas, J.~Cui, and B.~Ramabhadran,
\newblock ``Efficient knowledge distillation from an ensemble of teachers,''
\newblock in {\em Proc. INTERSPEECH}, 2017, pp. 3697--3701.

\bibitem{blending:scheme}
K.~J. Geras et~al.,
\newblock ``Blending lstms into cnns,''
\newblock in {\em Proc. ICLR Workshop}, 2016.

\bibitem{senior-et-al:scheme}
A.~Senior, H.~Sak, F.~C.~Quitry, T.~Sainath, K.~Rao, et~al.,
\newblock ``Acoustic modelling with cd-ctc-smbr lstm rnns,''
\newblock in {\em Proc. ASRU}, 2015, pp. 604--609.

\bibitem{takashima-et-al:scheme}
R.~Takashima, S.~Li, and H.~Kawai,
\newblock ``An investigation of a knowledge distillation method for ctc
  acoustic models,''
\newblock in {\em Proc. ICASSP}, 2018, pp. 5809--5813.

\bibitem{takashima-et-al2:scheme}
R.~Takashima, S.~Li, and H.~Kawai,
\newblock ``Investigation of sequence-level knowledge distillation methods for
  ctc acoustic models,''
\newblock in {\em Proc. ICASSP}, 2019, pp. 6156--6160.

\bibitem{sequence-level-kd}
Y.~Kim and A.~M. Rush,
\newblock ``Sequence-level knowledge distillation,''
\newblock in {\em Proc. EMNLP}, 2016.

\bibitem{kurata2-et-al:scheme}
G.~Kurata and K.~Audhkhasi,
\newblock ``Improved knowledge distillation from bi-directional to
  uni-directional lstm ctc for end-to-end speech recognition,''
\newblock in {\em Proc. SLT}, 2018, pp. 411--417.

\bibitem{kurata-et-al:scheme}
G.~Kurata and K.~Audhkhasi,
\newblock ``Guiding ctc posterior spike timings for improved posterior fusion
  and knowledge distillation,''
\newblock in {\em Proc. INTERSPEECH}, 2019, pp. 1616--1620.

\bibitem{tutornet:scheme}
J.~W. Yoon, H.~Lee, H.~Y. Kim, W.~I. Cho, and N.~S. Kim,
\newblock ``Tutornet: towards flexible knowledge distillation for end-to-end
  speech recognition,''
\newblock {\em IEEE/ACM Transactions on Audio, Speech, and Language
  Processing}, vol. 29, pp. 1626--1638, 2021.

\bibitem{li-et-al:scheme}
J.~Li, V.~Lavrukhin, B.~Ginsburg, R.~Leary, O.~Kuchaiev, J.~M. Cohen,
  H.~Nguyen, and R.~T. Gadde,
\newblock ``Jasper: An end-to-end convolutional neural acoustic model,''
\newblock {\em arXiv preprint arXiv:1904.03288}, 2019.

\bibitem{openseq2seq}
O.~Kuchaiev, B.~Ginsburg, I.~Gitman, V.~Lavrukhin, J.~Li, H.~Nguyen, C.~Case,
  and P.~Micikevicius,
\newblock ``Mixed-precision training for nlp and speech recognition with
  openseq2seq,''
\newblock {\em arXiv preprint arXiv:1805.10387}, 2018.

\bibitem{ginsburg-et-al:scheme}
B.~Ginsburg, P.~Castonguay, O.~Hrinchuk, O.~Kuchaiev, V.~Lavrukhin, R.~Leary,
  J.~Li, H.~Nguyen, and J.~M. Cohen,
\newblock ``Stochastic gradient methods with layer-wise adaptive moments for
  training of deep networks,''
\newblock {\em arXiv preprint arXiv:1905.11286}, 2019.

\bibitem{Heafield2011KenLMFA}
K.~Heafield,
\newblock ``Kenlm: faster and smaller language model queries,''
\newblock in {\em Proc. EMNLP}, 2011.

\end{thebibliography}

\end{document}